\documentclass[11pt]{article}
\usepackage{moriond,amssymb,epsfig}

\usepackage{subfigure}

\def\MyPhoto{\psfig{figure=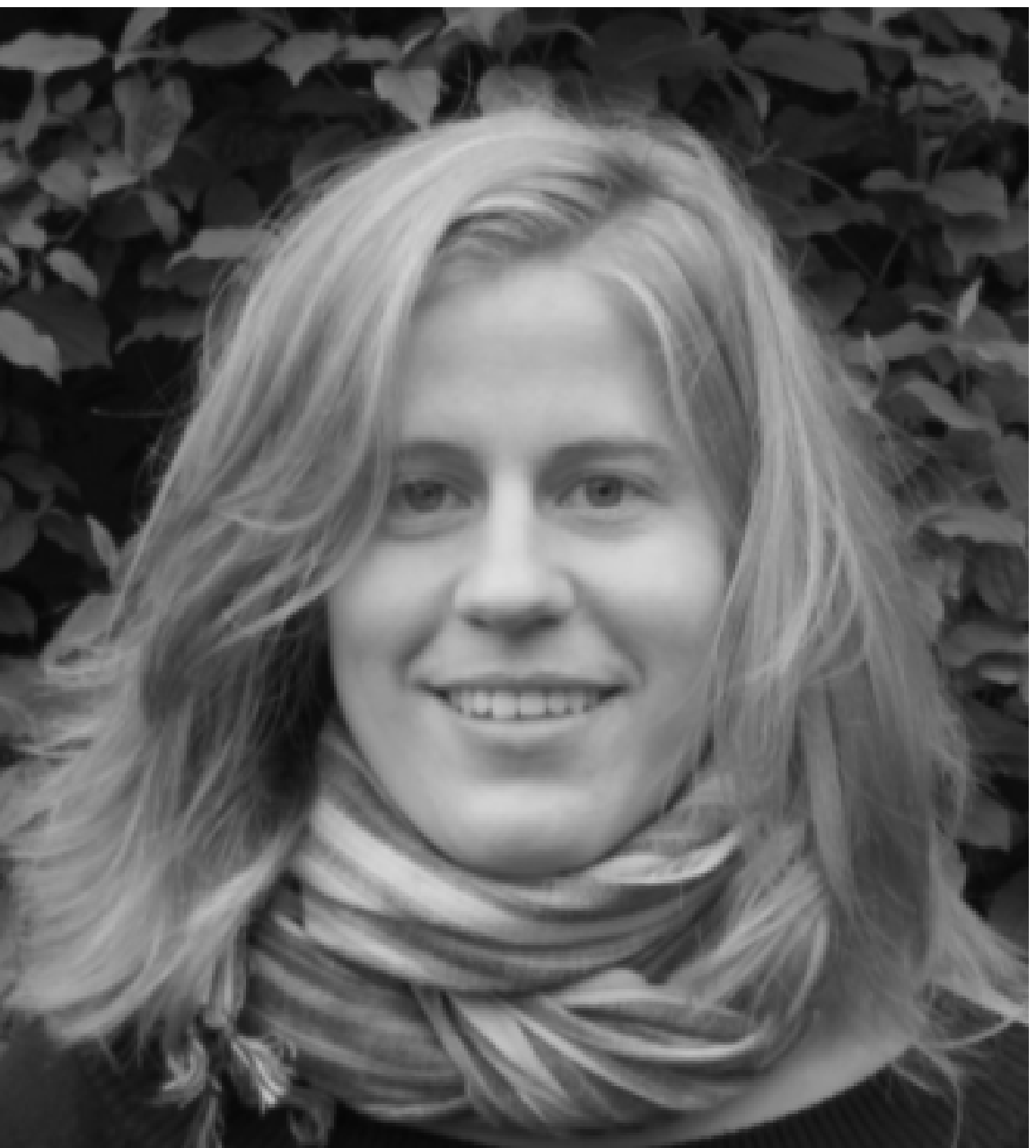,width=35mm}}

\def\PLB{{\em Phys. Lett.}  B}

\def\PRD{{\em Phys. Rev.} D}

\def\be{\begin{equation}}
\def\ee{\end{equation}}
\def\bea{\begin{eqnarray}}
\def\eea{\end{eqnarray}}

\newcommand{\unit}[1]{\; \mathrm{#1}}
\newcommand{\GeV}{\unit{GeV}}
\newcommand{\cm}{\unit{cm}}
\newcommand{\sqcm}{\cm^{2}}

\begin{document}
\vspace*{4cm}
\title{SCALAR DARK MATTER AND DAMA}

\author{ S. ANDREAS }

\address{Service de Physique Th\'eorique, Universit\'e Libre de
Bruxelles, B-1050 Brussels, Belgium\\
Institut f\"ur Theoretische Physik E, RWTH Aachen University,
D-52056 Aachen, Germany \\
\texttt{Sarah.Andreas@rwth-aachen.de}}

\maketitle\abstracts{A light scalar WIMP is studied in view of the
recent results of the DAMA collaboration. In a scenario where both
the WIMP's annihilation and its elastic scattering on nuclei occur
dominantly through Higgs exchange, a one-to-one relation between the
WIMP's relic density and its spin-independent direct detection rate
is established. The ratio of the relevant cross sections depends
only on the dark matter mass if the range allowed by the DAMA
results (m $<$ 10 GeV) is considered. We show that if such a light
scalar WIMP possesses a direct detection rate compatible with DAMA,
it naturally obtains a relic abundance in agreement with WMAP.
Indirect detection both with gammas from the Galactic centre and
neutrinos from the Sun opens possibilities to test this light dark
matter scenario.}

\section{Introduction}
The recent observation of an annual modulation in the nuclear recoil
rate by the DAMA collaboration~\cite{DAMA} can be reconciled with
null results of other direct detection searches~\cite{CDMSXenon} for
a spin-independent (SI) scattering cross section and dark matter
(DM) mass in the ranges
\begin{equation}
\label{SI} 3 \times 10^{-41} \sqcm \lesssim \sigma_p^{SI} \lesssim 5
\times 10^{-39} \sqcm \qquad \mathrm{and} \qquad 3 \GeV \lesssim
m_{DM} \lesssim 8 \GeV
\end{equation}
when taking into account the channeling
effect~\cite{Petriello:2008jj}. These results have already been
studied in various specific
models~\cite{OtherModels1,OtherModels2,OtherModelfrange,Andreas:2008xy}
and are discussed in the following in view of a light scalar
WIMP~\footnote{The WIMP relic density is determined by thermal
freeze-out assuming a mundane, radiation dominated expansion of the
universe. We use general assumptions on abundance and velocity of
the DM distribution in our neighbourhood.}. For this DM candidate we
also present the signature in indirect detection through gammas from
the Galactic centre (GC)~\cite{Andreas:2008xy} and neutrinos from
the Sun~\cite{Andreas:2009hj}.

\section{Light scalar Dark Matter}
\begin{figure}[h]
\begin{center}
\epsfig{file=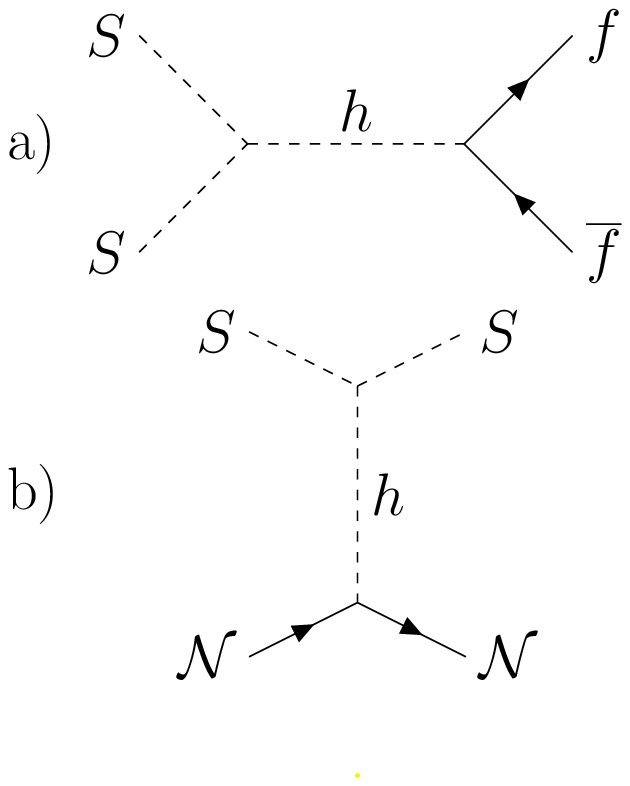,width=4cm} %
\hspace{1cm}
\epsfig{file=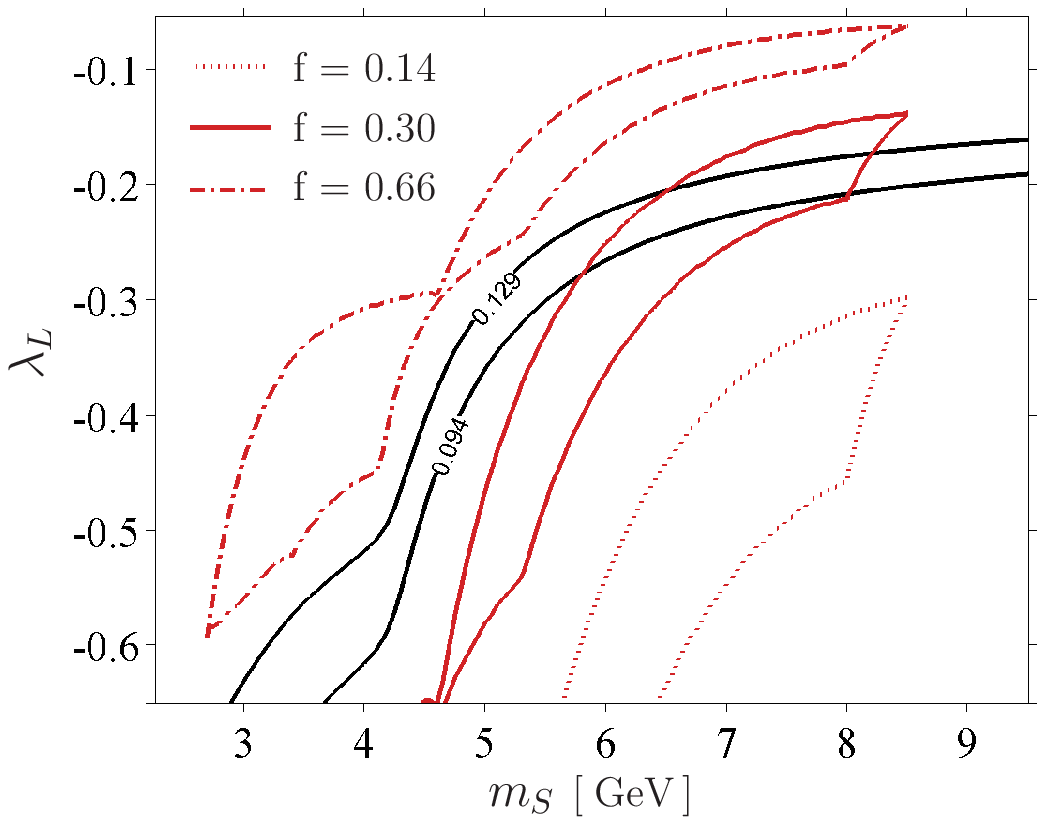,width=7cm}%
\caption{\textit{left}: Higgs exchange diagrams for the DM
annihilation (a) and scattering with a nucleon $\mathcal{N}$ (b).
$\qquad$ \textit{right}: The abundance is in agreement with WMAP
($0.094< \Omega_{DM} h^2 < 0.129$) between the black solid lines.
Direct detection constraints are met in the regions surrounded by
the red lines for $f=0.14$ (dotted lines), $f=0.30$ (solid lines)
and $f=0.66$ (dash-dotted lines). ($m_h=120 \GeV$)
\label{fig-Fey-DW}}
\end{center}
\end{figure}

Assuming only one Higgs boson, such a scenario is of interest since
for a light scalar DM candidate, a natural one-to-one relation
between the annihilation cross section and the one for SI scattering
on a nucleon~$\mathcal{N}$ arises as both processes occur in a
Higgs-channel and connects the DM abundance to the direct detection
rate. The sole possible tree level annihilation process
(figure~\ref{fig-Fey-DW}a) is through a Higgs boson in a s-channel
into a pair of fermions, among which only $b \bar{b}$, $c \bar{c}$
and $\tau \bar{\tau}$ are relevant, since all other SM fermions have
small Yukawa couplings~\footnote{Annihilation through the SM $Z$
boson is excluded because the DM would contribute to the $Z$
invisible width~\cite{PDG}.}. The SI elastic scattering occurs
exclusively through a Higgs in the t-channel
(figure~\ref{fig-Fey-DW}b) and is induced by the same DM-Higgs
coupling.

The scalar DM candidate is introduced in the simplest way as a real
scalar singlet~$S$, odd under a $Z_2$ symmetry, by adding the four
following renormalizable terms to the SM lagrangian:
\begin{equation}
\label{lag} {\cal L} \owns \frac{1}{2}\partial^\mu S \partial_\mu
S-\frac{1}{2}\mu^2_S \,S^2 -\frac{\lambda_S}{4} S^4 -\lambda_L\,
H^\dagger H\, S^2
\end{equation}
with the Higgs doublet $H=(h^+ \, (h+iG_0)/\sqrt{2})^T$. The mass of
$S$ is thus given by $m_S^2=\mu^2_S+\lambda_L \mathrm{v}^2$ where
$\mathrm{v}=246 \GeV$. In this model, the sole coupling which allows
$S$ to annihilate into SM particles and to interact with nucleons is
$\lambda_L$. For the annihilation and the elastic scattering cross
section (normalized to one nucleon), we obtain
\begin{eqnarray}
\sigma (S S \rightarrow \bar{f} f) v_{rel} &=& n_c \,
\frac{\lambda_L^2}{\pi} \frac{m_f^2 }{m_h^4 m_S^3}
(m_S^2-m_f^2)^{3/2} \label{sigmaSannih}\\
\sigma (S \mathcal{N} \rightarrow S \mathcal{N}) &=&
\frac{\lambda_L^2}{\pi} \frac{m_{\mathcal{N}}^4}{m^4_h \ (m_S +
m_{\mathcal{N}})^2} f^2, \label{sigmaSdirectdet}
\end{eqnarray}
where ${n_c}=3(1)$ for quarks (leptons) and $v_{rel}=(s-4
m_S^2)^{1/2}/m_S$ is the centre of mass relative velocity between
both $S$. The factor $f$ parametrizes the Higgs to nucleons coupling
from the trace anomaly, $f m_{\mathcal{N}}\equiv \langle
{\mathcal{N}}| \sum_q m_q
\bar{q}q|{\mathcal{N}}\rangle=g_{h{\mathcal{N}\mathcal{N}}}
\mathrm{v}$. Reflecting the uncertainty in $f$ we consider the range
$0.14 < f < 0.66$ with $f=0.30$ as central
value~\cite{OtherModelfrange}. As for the Yukawa
couplings~\footnote{We neglect the effects of the running of the
Yukawa couplings which are expected to be quite moderate.},
$Y_i=\sqrt{2}m_i/ \mathrm{v}$, we consider the pole masses $m_b=4.23
\GeV$, $m_c=1.2 \GeV$ and $m_\tau=1.77 \GeV$.

\section{DAMA and WMAP}
In function of the two free parameters $m_S$ and $\lambda_L$ and for
a Higgs mass $m_h = 120 \GeV$, we check whether or not agreement
with both experiments can be obtained. For the parameter space
between the two black lines (figure~\ref{fig-Fey-DW}), the relic
density, computed with \textsc{micrOMEGAs}~\cite{Belanger:2006is},
with respect to the critical density lies within the WMAP density
range $0.094< \Omega_{DM} h^2 < 0.129$~\cite{WMAP}. In the red
regions $\sigma^{SI}_p$ and $m_S$ are in agreement with DAMA and
allowed by other direct detection
experiments~\cite{Petriello:2008jj} (taking into account the
channeling effect~\footnote{There is no region allowed by all
experiments if no channeling effect is assumed.}). We find that the
regions of $m_S$ and $\lambda_L$ which are consistent with WMAP and
direct detection constraints nicely overlap~\footnote{A DM mass in
the relevant range demands some fine tuning, which is in our opinion
not unbearable nor surprising, given the minimal number of
parameters of the model. To obtain the quite large SI cross section
needed to fit the DAMA data a large, albeit still perturbative
coupling is required.}. For the central value $f=0.30$, the overlap
ranges over $m_S\approx 6 - 8 \GeV$ while for $0.14<f<0.66$ regions
overlap for $3.5 \GeV < m_S < 8.4 \GeV$. For $f$ smaller than $0.20$
no overlap exists.

Those results are shown for $m_h=120$ GeV but agreement may be
obtained for other Higgs masses provided the ratio $\lambda_L/m_h^2$
is kept fixed, typically at a value $\lambda_L/m_h^2
\simeq10^{-5}$~GeV$^{-2}$. To keep the result perturbative
($\lambda_L \lesssim 2 \pi$) we need that $m_h \lesssim 800 \GeV$.

\section{Indirect Detection and LHC}
For the parameter range of interest, we make predictions regarding
possible indirect detection from DM annihilation through gamma rays
from the GC and neutrinos from the Sun.

\begin{figure}[h]
\begin{center}
\epsfig{file=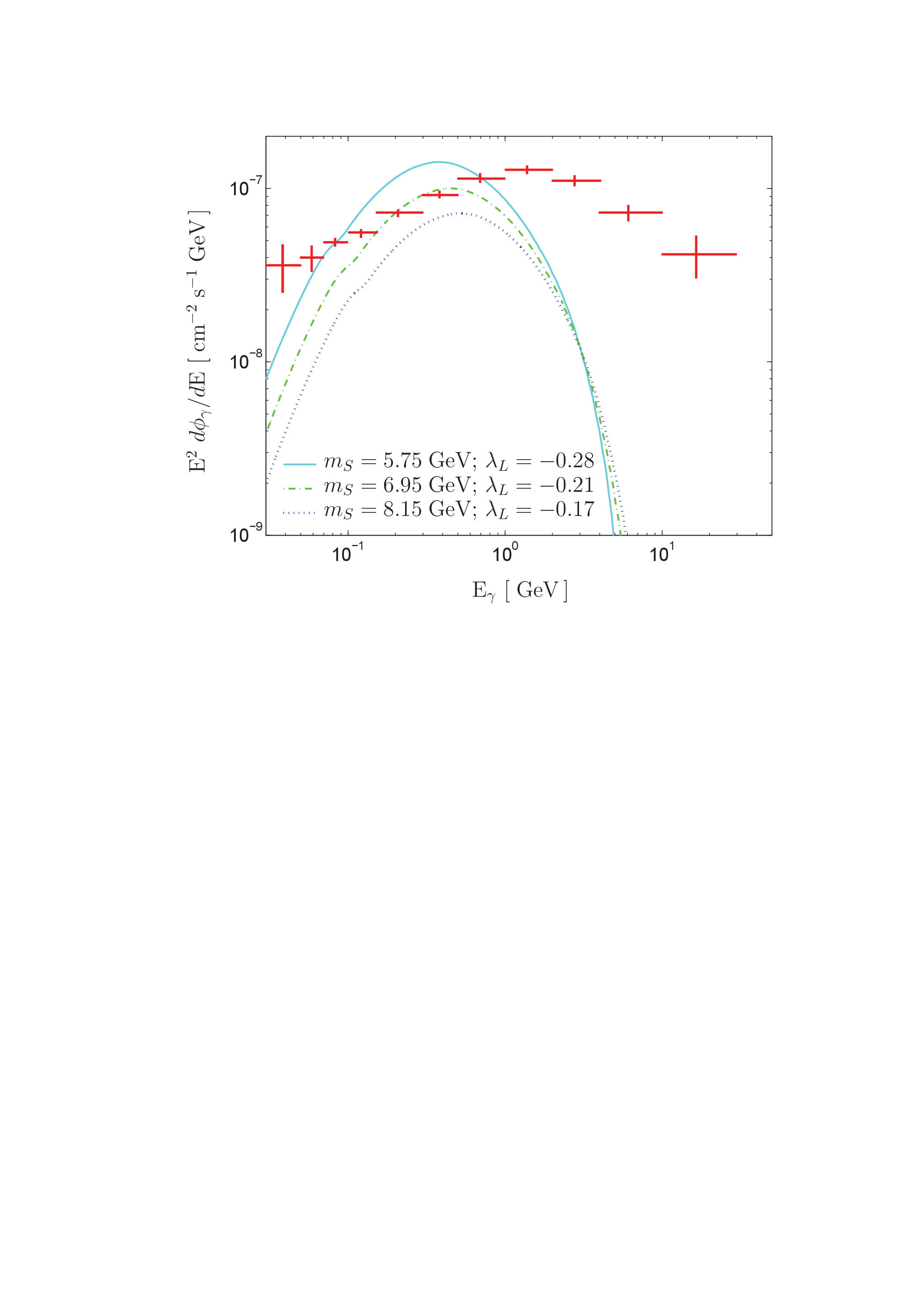,width=6.5cm}%
\hspace{0.3cm}
\epsfig{file=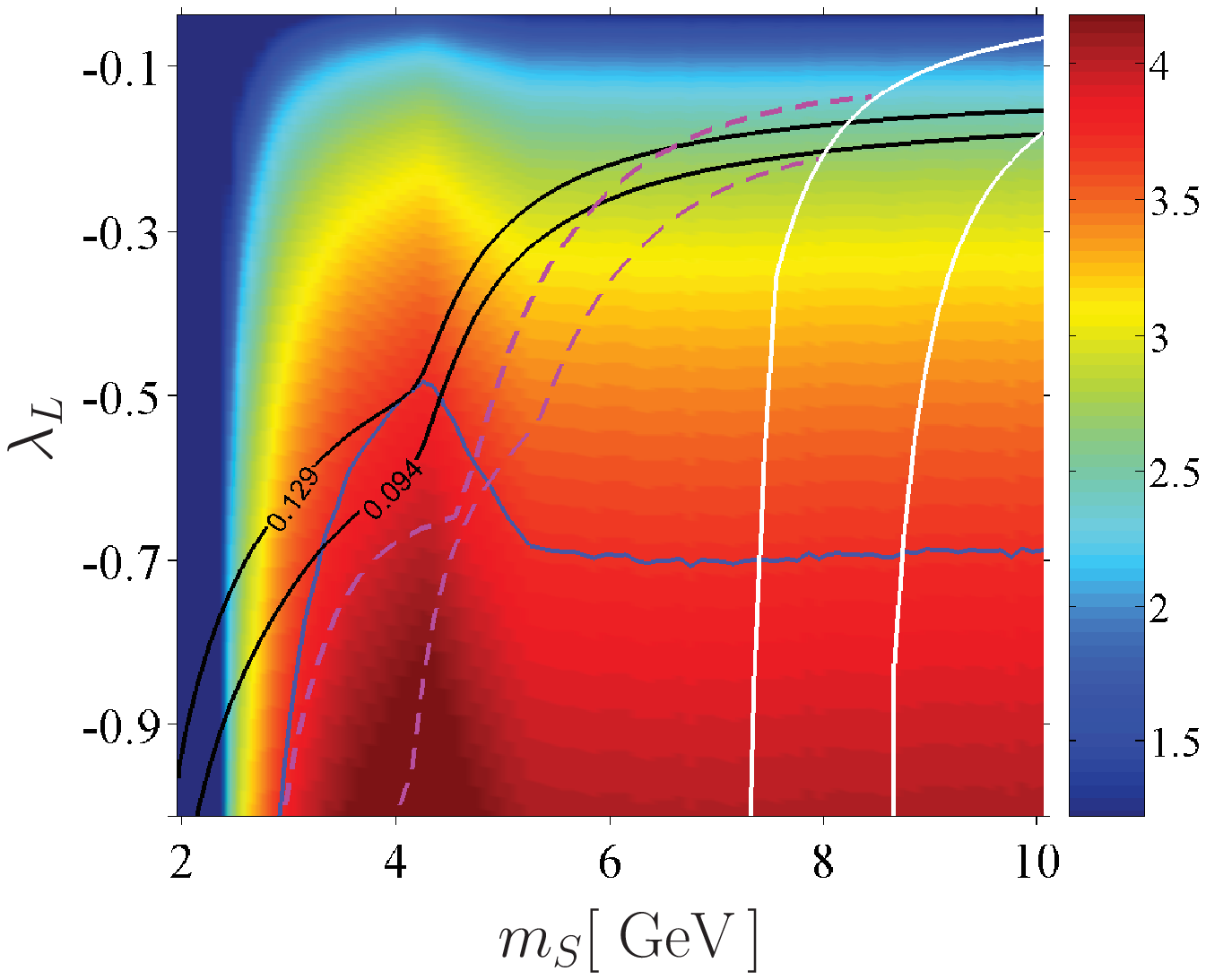,width=6.5cm}%
\caption{\textit{left}: Flux of gammas from DM-annihilation at the
GC for three examples of $m_S$, $\lambda_L$ consistent with DAMA,
compared to EGRET data (red crosses). \textit{right}: Expected
neutrino induced muon flux ($\log_{10} \phi_\mu$) from the Sun, for
2 GeV neutrino energy threshold, together with estimated
Super-Kamiokande sensitivity conservatively extrapolated to light
WIMPs (blue line, sensitive below), XENON and CDMS exclusion limits
(white lines, from left to right), DAMA (magenta dashed lines) and
WMAP (black lines) regions. \label{fig-GC-Sun}}
\end{center}
\vspace{-0.35cm}
\end{figure}

The gamma fluxes from the GC for a NFW profile are shown in
figure~\ref{fig-GC-Sun} for three parameter sets consistent with
DAMA and WMAP and $m_h = 120 \GeV$. Since those gammas have an
energy in the range of the EGRET (and the forthcoming FERMI/GLAST)
data we give for comparison the flux seen by
EGRET~\cite{Hunger:1997we}. The predicted flux is of the same order
of magnitude and may even be larger than the one observed. We have
however refrained from putting constraints on  model parameters,
given the large uncertainties on the DM density at the GC.

Dark matter which has been captured in the Sun~\cite{Jungman:1995df}
annihilates and produces neutrinos that can be observed at the Earth
after converting into muons close to the detector volume. The
expected flux of neutrino induced muons from the Sun for $m_h = 120
\GeV$ is presented in figure~\ref{fig-GC-Sun} together with the WMAP
and DAMA regions from figure~\ref{fig-Fey-DW} as well as the CDMS
and XENON limits. With our horizontally from
higher~\cite{Desai:2004pq} to lower DM masses conservatively
extrapolated Super-Kamiokande sensitivity~\footnote{The low
energetic neutrinos from light DM might be observed with
Super-Kamiokande if its sensitivity is extended down to 2 GeV by
including stopped, partially contained or fully contained
muons~\cite{Feng:2008qn}.} (blue line) it can be seen that
Super-Kamiokande is potentially able to test the light scalar DM and
a part of the DAMA allowed region ($3 - 4.8
\GeV$)~\cite{Andreas:2009hj}.

Additionally, in this framework, the Higgs boson is predicted to be
basically invisible at LHC for $m_h=120 \GeV$. The large coupling to
the Higgs leads to its large decay rate into a pair of scalar WIMPs,
e.g. for $m_S=7 \GeV$, $BR_{h \rightarrow SS}=99.5\,\%$ for
$\lambda_L=-0.2$, $m_h = 120 \GeV$ and $BR_{h \rightarrow SS}\simeq
70\,\%$ for $\lambda_L=-0.55$, $m_h=200 \GeV$~\cite{Andreas:2008xy}.

\section{Conclusions}
Light scalar DM has a one-to-one relation between SI direct
detection rate and relic abundance since both processes occur in a
Higgs channel. We show that this model can at the same time give the
correct WMAP abundance and account for the DAMA results without
contradicting other direct searches. The presented signatures in
indirect detection show potential to further test this model. Gamma
rays from the GC might be in reach of the upcoming FERMI/GLAST
satellite and Super-Kamiokande might set constraints through
neutrinos from the Sun. A striking consequence for LHC Higgs
searches is the severe reduction of the visible branching
ratio.\footnote{The results discussed here apply for any scalar DM
model (like the inert doublet or Higgs portal models) for which
annihilation and SI scattering would be dominated by the diagrams of
figure~\ref{fig-Fey-DW}a,b. For a fermionic DM candidate we find
that other channels must be present in order to match the WMAP and
DAMA observations~\cite{Andreas:2008xy}.}

\section*{Acknowledgments}
The work presented here was done in collaboration with Michel H. G.
Tytgat and Thomas Hambye. We thank Jean-Marie Fr\`ere for
stimulating discussions. Preprint ULB-TH/09-09.

\end{document}